\documentclass[aps,twocolumn,groupedaddress,amssymb]{revtex4}

\usepackage{latexsym}
\usepackage{amsbsy}
\usepackage{amsmath}
\usepackage{amssymb}
\usepackage{graphicx}
\usepackage{easyReview}
\usepackage{xcolor}
\usepackage{hyperref}
\hypersetup{
    colorlinks=true,
    linkcolor=black,
    urlcolor=blue,
    citecolor=blue
}
\usepackage[capitalise]{cleveref}

\begin{document}

\title{Occurrence of chemically tuned spin-texture controlled large intrinsic anomalous Hall effect in epitaxial $Mn_{3+x}Pt_{1-x}$ thin Films}


\author{Indraneel Sinha\textsuperscript{1}, Saurav Sachin\textsuperscript{1},
Shreyashi Sinha\textsuperscript{1}, Roumita Roy\textsuperscript{2,3}, Sudipta Kanungo\textsuperscript{3}\footnote{Contact author: sudipta@iitgoa.ac.in}, Sujit Manna\textsuperscript{1}\footnote{Contact author: smanna@physics.iitd.ac.in} }

\affiliation{\textsuperscript{1}Department of Physics, {Indian Institute of Technology Delhi}, Hauz Khas, New Delhi 110016, India}
\affiliation{\textsuperscript{2}Consiglio Nazionale delle Ricerche, CNR-SPIN, c/o Universitá “G. D’ Annunzio”, 66100 Chieti, Italy.}
\affiliation{\textsuperscript{3}School of Physical Sciences, {Indian Institute of Technology Goa}, Goa 403401, India}

\begin{abstract}
  Achieving atomically flat and stoichiometric films of chiral antiferromagnets (AFM) with two-dimensional kagome spin lattice structures are crucial for integrating these materials in both established and emerging antiferromagnetic spintronic devices. We report a systematic study of growth and anomalous Hall effect in (111)-oriented non-collinear AFM $Mn_{3+x}Pt_{1-x}$ films with varying compositions, for x = 0.09, 0.17, 0.28. Under optimized growth conditions, we obtain stoichiometric and atomically flat epitaxial $Mn_{3}Pt$(111) films on Si(100) substrate, as evidenced by X-ray reflectivity and scanning probe microscopy. The magnetization measurement showed that epitaxial strain can induce a magnetic phase transition from an incommensurate spin state ($T_2$) at x = 0.09 to a triangular all-in/all-out AFM spin order ($T_1$) at x = 0.17, 0.28. The change in magnetic ground state is evident in the transport characteristics, as the ($T_1$) state shows a robust intrinsic anomalous Hall effect (AHE) persisting till room temperature, in contrast to the ($T_2$) state where AHE is negligible. Our studies reveal a hole-dominated conductance with room temperature anomalous Hall conductivity (AHC) ranging from 5 to 16 $\Omega^{-1}cm^{-1}$ for x = 0.17 and 0.28 respectively. A scaling law is established, indicating that Hall resistivity is primarily governed by the intrinsic non-vanishing Berry curvature. The experimental observation corroborates the electronic structure calculations, which predicts the massless Dirac states near Fermi level in the bulk band structure, attributed to the presence of nonsymorphic glide symmetry. Additionally, we showed that chemical tuning via Mn doping can stabilize the required T$_1$ non-collinear AFM structure which enhance the topology driven intrinsic AHE. 

\end{abstract}

\maketitle


\section{\label{sec:level1}Introduction}
\begin{figure*}
    \centering
    \includegraphics[width=01.0\textwidth]{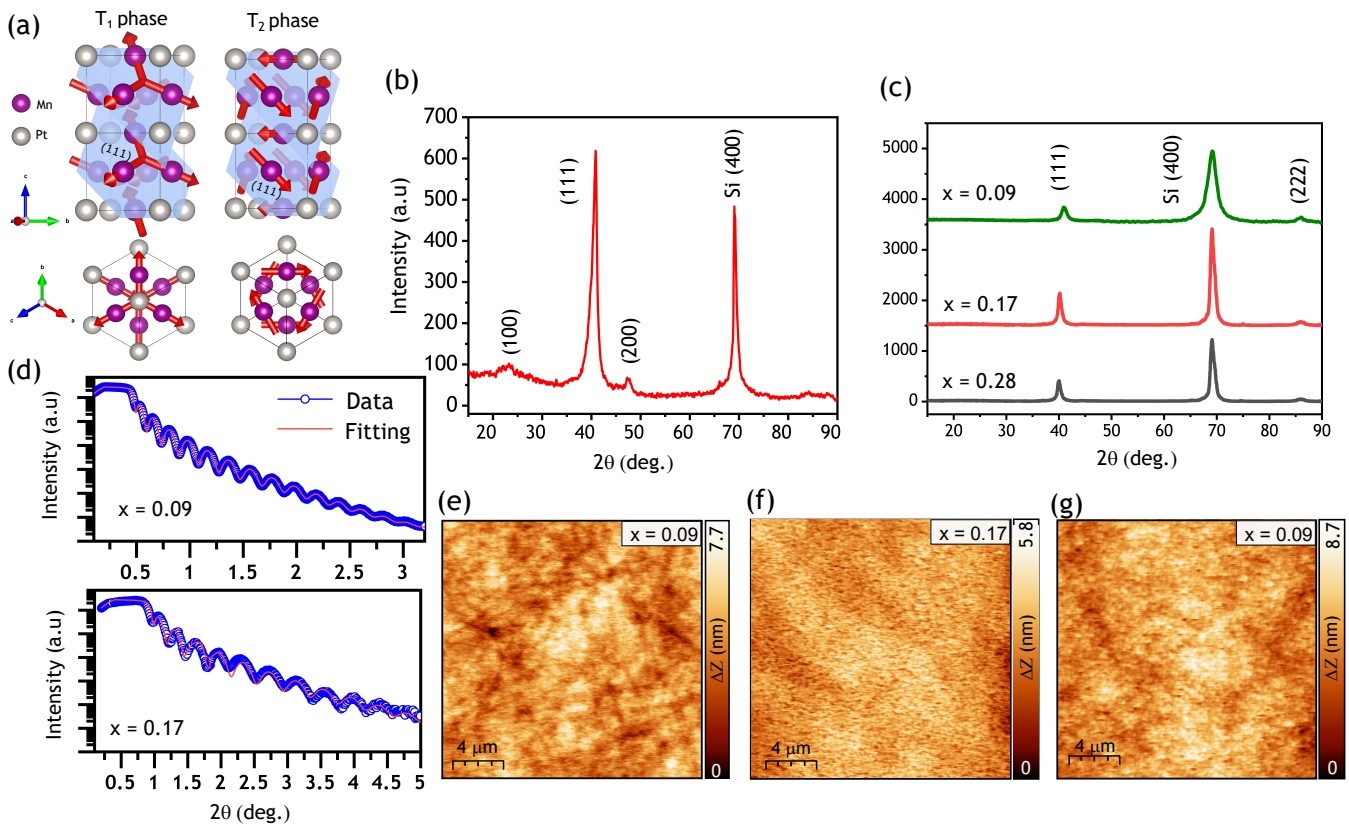} 
    \caption{ Crystal structure determination using X-ray diffraction. (a) Schematic of two possible spin states of cubic Mn$_{3}$Pt: $T{1}$ (left) represents the topological phase with Mn spins in a tail-to-tail configuration, while $T_{2}$ (right) corresponds to the non-topological phase with a head-to-tail configuration. The Kagome lattice lies in the (111) plane for both spin configurations. (b) Out-of-plane XRD diffraction profile measured on polycrystalline Mn$_{3}$Pt films grown at 400 °C on Si(100)substrates. The scan shows multiple peaks corresponding to the (100) and (111) planes, with a preferential growth along the (111) direction. (c) Out-of-plane XRD pattern of epitaxial Mn$_{3+x}$Pt$_{1-x}$ thin films with different compositions ($x = 0.09, 0.17, 0.28$) deposited at 600 °C. The (111) peak position shifts toward lower angles in Mn-rich samples. (d) X-ray reflectivity data measured on Mn$_{3+x}$Pt$_{1-x}$ thin films for $x=0.09$ and $x=0.17$. Blue balls represent experimental data points, and the solid red line shows the best fit to the data. (e-g) Atomic force microscopy images showing the surface topography of Mn$_{3+x}$Pt$_{1-x}$ thin films for $x = 0.09, 0.17, 0.28$, respectively.}
    \label{fig:break1}
\end{figure*}

Non-collinear antiferromagnets with Kagome spin structures have gained significant interest due to their chirality-driven novel phenomena, offering promising potential for high-density and non-volatile spintronic devices \cite{pal2022setting,hazra2023generation,jeon2021long}. This two-dimensional network of corner-sharing triangles serves as a rich platform for exploring the interplay between topology, electron correlations, and magnetism \cite{zhang2022progress,zhang2017strong,yang2017topological}. The Kagome lattices have been observed in several layered systems, like $AV_3Sb_5$ (A = K, Rb, Cs), perovskite oxides [(Y, Lu)MnO\textsubscript{3}, SmFeO\textsubscript{3}] or ferromagnetic alloys like Co$_3$Sn$_2$S$_2$, Fe$_3$Sn$_2$, and Fe$_5$Sn$_3$ \cite{ortiz2021superconductivity,okamura2020giant,ren2022plethora,li2020large,xu2022observation}.  In particular, the binary antiferromagnetic alloys $Mn_{3}Y$ (Y = Sn, Ge, Pt, Rh) have remained in the frontier for their intriguing topological signatures like the anomalous Hall effect (AHE), anomalous Nernst effect (ANE), and magneto-optical Kerr effect \cite{chen2021anomalous,nayak2016large,ikhlas2017large,higo2018large}. This family of compounds is known to crystallize in either face-centered cubic (fcc) (Y = Ir, Pt, and Sb) or  hexagonal close-packed (hcp) for Y = (Ga, Ge, Sn). The Kagome lattices are stacked along the (111) and (0001) for cubic and hexagonal systems, respectively \cite{mccoombs2023impact,yang2017topological}. The magnetic ground state consists of a geometrically frustrated lattice with Mn atoms  arranged in an inverse triangular spin configuration stabilized by the Dzyaloshinskii-Moriya interaction \cite{nagamiya1982triangular,tomiyoshi1982magnetic}. The large AHE in such a small magnetic background is explainable from the intrinsic contribution of the Berry curvature in the momentum space. In cubic systems, the Berry curvature is preserved by the interplay between time-reversal ($\mathcal{T}$) and mirror ($\mathcal{M}$) symmetries, represented as $\mathcal{TM}$. The application of an external magnetic field perpendicular to the plane breaks this symmetry ($\mathcal{TM}$), leading to a finite Berry curvature in momentum space \cite{chen2014anomalous}.
The Berry curvature-induced large anomalous Hall effect (AHE) under an external magnetic field has been mostly studied in the hexagonal systems, particularly $Mn_{3}Ge$ and $Mn_{3}Sn$ \cite{nayak2016large,taylor2020anomalous,takeuchi2024magnetic}. The hexagonal members of the family are imposed with lower crystalline symmetry, leading to effective and faster switching at moderate fields. The study of the cubic phase is also limited due to the complicated  thermodynamics of several compatible phases. Recently, Mukherjee et al. observed large AHE at room temperature in polycrystalline Mn$_3$Pt thin films with a reversal in carrier type \cite{mukherjee2021sign}. In 2018, Liu et al. synthesized high-quality (001)-oriented Mn$_3$Pt films, exhibiting an anomalous Hall conductivity (AHC) of approximately 98 $\Omega$ cm$^{-1}$ \cite{liu2018electrical}. Later, An et al. demonstrated a structural transition from $L_{10}$ MnPt to $L_{12}$ Mn$_3$Pt through compositional variation \cite{an2020structure}.
Zhao et al. studied the effects of strain on the magnetic and transport properties by varying the thickness of  $Mn_{3}Pt$ thin films \cite{zhao2022strain}.
Among other cubic systems, epitaxial $Mn_{3}Ir$ thin films grown using sputtering is reported with an AHC of 40 $\Omega^{-1}$ cm$^{-1}$  at 300 K \cite{iwaki2020large}.
$Mn_{3}Pt$ has four potential magnetic ground states, two manifest non-collinearity in their spin structures \cite{kren1968magnetic}. \hyperref[fig:break1]{\textcolor{blue}{Fig \ref*{fig:break1}(a)}} shows two distinct AFM spin configurations labeled as $T_{1}$ and $T_{2}$ respectively. The tilting of the spins plays a key role in shaping the non-vanishing Berry curvature \cite{chen2014anomalous,bai2021control}. $T_{2}$ can be viewed as a head-to-tail spin configuration with Mn spins pointing along the [110] direction. $T_{2}$ phase was previously observed in $Mn_{3}Sn$ single crystals as an intermediate non-topological state between the spin glass and chiral states \cite{deng2021effect,sung2018magnetic}. Recently, Liu et al. reported switching between the D phase ($T_{1}$) to a collinear state (F phase) using piezoelectric-mediated strain tuning \cite{liu2018electrical}.  Magnetic phase transition triggered via strain and compositional change are widely observed in binary alloys \cite{feng2019electric,lu2011magnetic,liu2018electrical}. Experimental studies on cubic nitrides $Mn_3MN$ (M= Ga, Sn, Ag, Zn, Rh, Pt) also demonstrate a  transition between the two triangular configurations ($T_{2}$ \& $T_{1}$), depending  on the sign of magnetic anisotropy \cite{fruchart1978magnetic}. A similar spin transition from $T_{2}$ to $T_{1}$  has been observed in in Cr-doped $Mn_3GaN$, driven by magnetovolume effect\cite{wang2024new}. 

In this study, we investigate the transition between two antiferromagnetic (AFM) states $T_{1}$ and $T_{2}$ through compositional variation in the system $Mn_{3+x}Pt_{1-x}$. We show the detailed synthesis of highly ordered $Mn_{3}Pt$ (111) thin films on Si(100) substrates, exhibiting both $T_{1}$ and $T_{2}$ spin configurations. By systematically varying the Mn:Pt composition ($x = 0.09, 0.17, 0.28$), we explore the corresponding magnetic and magnetotransport properties of the material. 
\begin{figure*}
    \centering
    \includegraphics[width=0.70\textwidth]{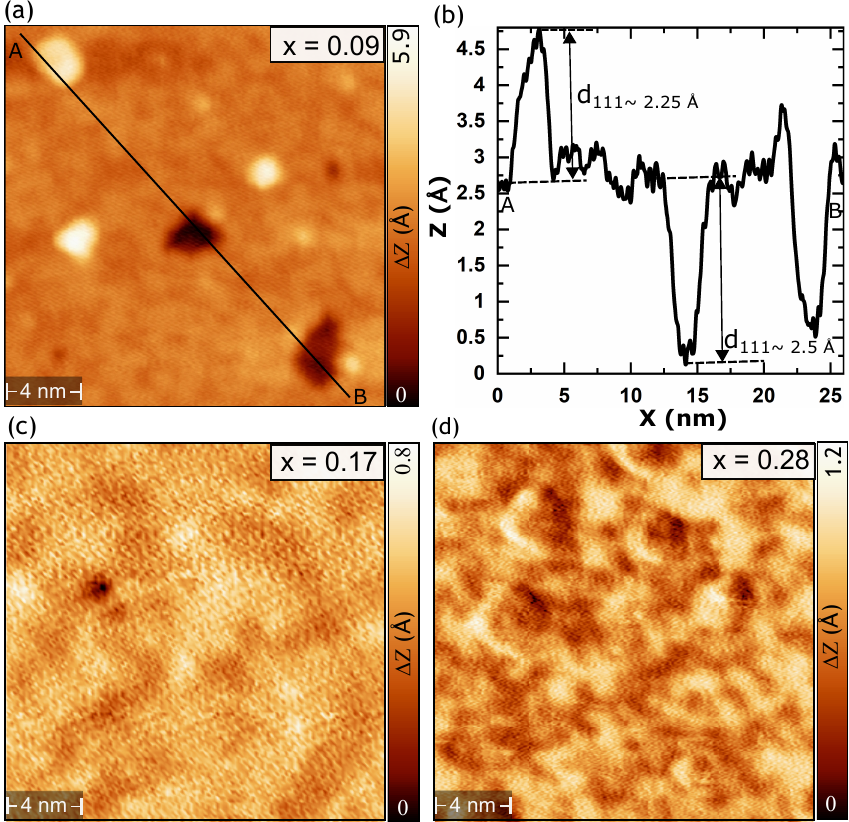} 
    \caption{Constant current STM topography images ($20 \times 20 \, \text{nm}^2$) acquired on the  $Mn_{3+x}Pt_{1-x}$ thin films for  (a) x = 0.09, (c) 0.17 and (d) 0.28. (a) Topography of atomically flat  $Mn_{3}Pt$ surface consisting of small islands and shallow holes originating from the different atomic layers terminations along the (111) axis. The bright islands refer to the crystallites of the topmost unfinished layer and the depression area represent the under layers. (b) The measured height profile along the black line (A-B) reveals an interlayer distance of approximately $2.25,\text{\AA}$, confirming a layer-by-layer growth with a (111) termination. (c-d) STM images measured on  Mn-rich sample grown with (x = 0.17, 0.28) show similar atomically flat surface termination. The images are recorded at constant current ($I_{s} = 1-1.2~nA$) mode with sample $V_{bias} = 150-200~mV$. }
    \label{fig:break2}
\end{figure*}
 
 \section{\label{sec:experimental_details}Experimental details}
$Mn_{3}Pt$ thin films were synthesized on Si(100) substrate in ultra-high vacuum (UHV) magnetron sputtering chamber with a base pressure below 2 x 10$^{-8}$. We prepared $Mn_{3+x}Pt_{1-x}$ films with varying compositions(x = 0.09, 0.17 and 0.28) using co-sputtering of Mn (99.9\%) and Pt (99.999\%) targets, powered by direct current (DC) and radio
frequency (RF) sources respectively. The Si(100) substrates were prepared by standard etching method followed by chemical cleaning. Prior to the deposition, the substrate are degassed in UHV, followed by repeated cycles of in-situ annealing up to 600°C. Thin films were grown in an argon environment with a pressure close to $2.0 \times 10^{-3} \, \text{mbar}$. To understand  the optimal growth conditions for both $T_{1}$ and $T_{2}$ phases, the  sputtering power of Pt is fixed at 25 W while the Mn sputtering power is varied between 40 W to 60 W in steps of 10W. Structural characterization was performed using X-ray diffraction (XRD) with a PANalytical X'Pert diffractometer with Cu-K$\alpha$ source ($\lambda = 1.5418 \, \text{\AA}$). The thickness of the films was calibrated using X-ray reflectivity (XRR) measurements to establish a standard between thickness and growth time. The surface morphology was analyzed using atomic force microscopy (AFM) (Oxford Instruments Asylum Research, MFP-3D system). Scanning tunneling microscopy measurements were carried out at room-temperature using Quaza STM, equipped with a home-built integrated active vibration cancellation system \cite{sinha2025magnetic}. The probing tips included electro-chemically etched PtIr and W tips subjected to annealing and subsequent field emission. The elemental composition of  Mn$_{3}$Pt films were determined using energy-dispersive X-ray analysis (Hitachi Table Top TM3000). The magnetization of the Mn$_{3}$Pt films were measured using Quantum Design magnetic property measurement system (MPMS-3). Temperature and field-dependent magnetotransport measurements are performed in a standard four-probe geometry using a cryogen-free Physical Property Measurement System (PPMS).

 \begin{figure*}
    \centering
    \includegraphics[width=1.0\textwidth]{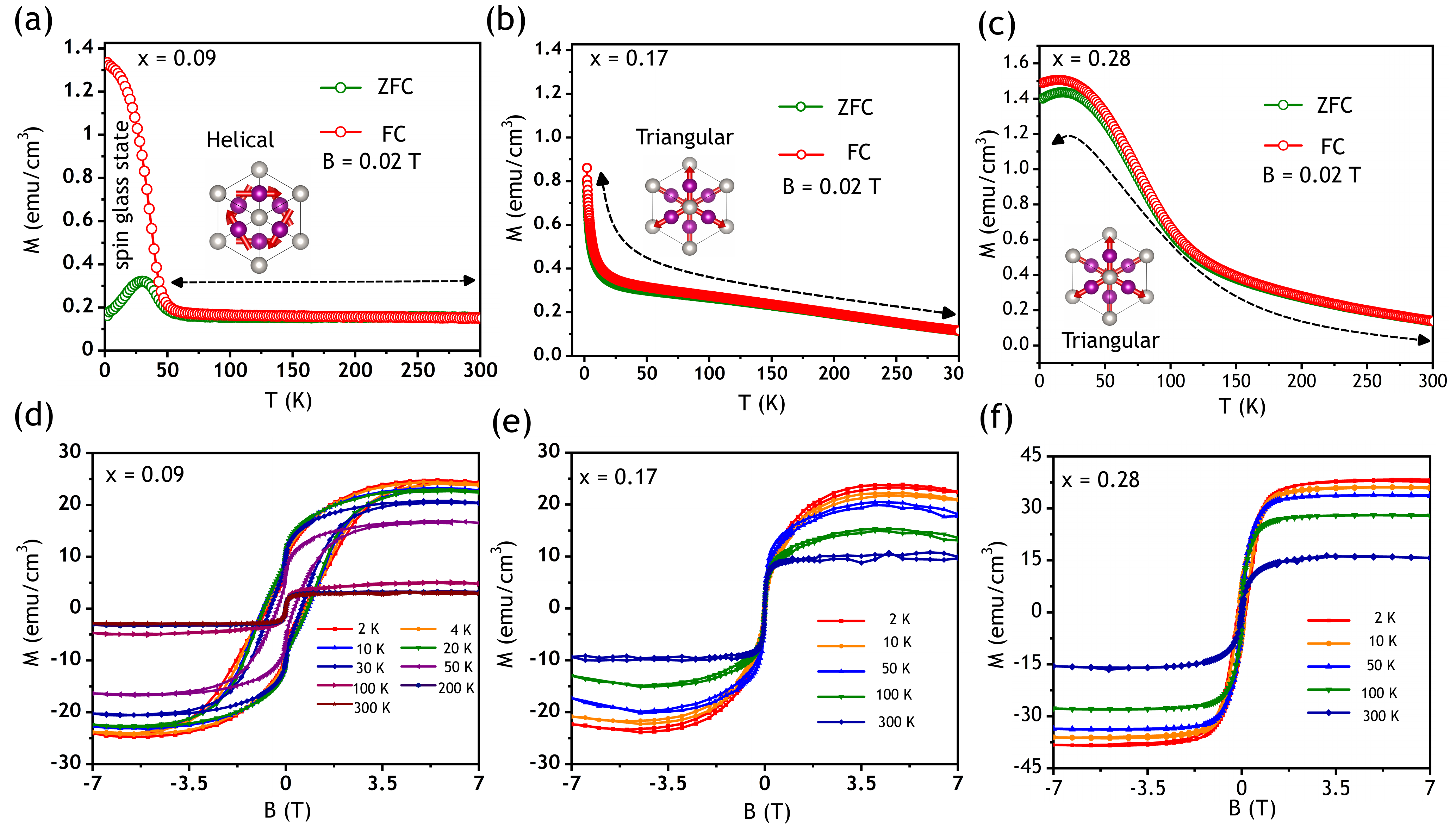} 
    \caption{ Out-of-plane magnetization measurements of  Mn$_{3+x}$Pt$_{1-x}$ thin films grown with x = (0.09, 0.17, 0.28). Temperature-dependent magnetization M(T) measured in zero-field-cooled (ZFC) and field-cooled (FC) modes for (a) x = 0.09, (b) x = 0.17, and (c) x = 0.28. The magnetic phase transition in (a) at 30 K corresponds to the transition from  a metastable spin glass phase to helical $T_{2}$ state. The magnetization curves in (b) and (c) follow reversible paths during ZFC and FC, indicating the stability of the triangular spin order ($T_{1}$) between 2-300 K. Field-dependent magnetization isotherms were measured between 2-300 K for (d) x = 0.09, (e) x = 0.17, and (f) x = 0.28. The large coercive field in (d) originates from the out-of-plane spin canting in the spin glass state. At 300 K, a vanishingly small ($M_{S}$ =~$2.5~emu/cm^{3}$) originates from the helical $T_{2}$ state. (e-f) Isotherms exhibit negligible coercive fields as the triangular spin structure is preserved over the entire temperature range. The observed value of $M_{S}$ at 300 K is around $10~emu/cm^{3}$ (0.17) and $15~emu/cm^{3}$ (0.28), attributed to the uncompensated Mn spins occupying Pt sites.  }
    \label{fig:break3}
\end{figure*}

\section{\label{sec:level2}Calculation Details}

The DFT calculations were conducted utilizing a plane-wave based basis set with a 500 eV cut-off within a pseudopotential framework employing the Perdew-Burke-Ernzerhof (PBE) \cite{perdew1996generalized,perdew1998perdew} exchange-correlation functional as implemented in the Vienna \textit{ab initio} simulation package (VASP) \cite{kresse1993ab,kresse1996efficient}. To capture the electron-electron Coulomb correlations within the Ir-5\textit{d} states, we employed an onsite Hubbard $U$ ($U_{eff} = U - J_H$) \cite{anisimov1993density,dudarev1998electron}. The spin-orbit coupling (SOC) effect has been incorporated in the calculations through relativistic corrections to the original Hamiltonian \cite{hobbs2000fully,steiner2016calculation}. We used a $6\times6\times8$ k-mesh in the hexagonal Brillouin zone (BZ) for the self-consistent calculations. The structural optimization was performed by relaxing the atomic positions towards equilibrium until the Hellmann-Feynman force became less than 0.001 eV/$\AA$, keeping the lattice unaltered.

\section{\label{sec:level2}Results and Discussions}
The crystal structure of the films with varying Mn composition is investigated using out-of-plane x-ray diffraction patterns. In our experiment, $Mn_{3}Pt$ films grown between 350 $^{\circ}$ C-500 $^{\circ}$ C are  polycrystalline in nature with dominant contributions from  (111) and (100) sets of planes \cite{mukherjee2021sign}. The relative intensities of the (100) and (111) planes in {\textcolor{blue}{Fig \ref*{fig:break1}(b)}} indicate that films are predominantly oriented to the (111) axis. In order to achieve epitaxial growth, the growth temperature is increased to 600 $^{\circ}$ C. The XRD profile in {\textcolor{blue}{Fig \ref*{fig:break1}(c)}} confirms the growth of highly oriented $Mn_{3}Pt$ thin films with reflections from (111) and (222) planes. The strong contribution from the Kagome plane (111) is also  an indication of long-range chemical ordering \cite{liu2018electrical}. The effect of epitaxial strain can be evidenced from the (111) peak position, which shifts to lower diffraction angles with increasing Mn concentration \cite{olayiwola2023room}. The (111) peak position indicates that the films grown with $x = 0.09$ composition are fully relaxed with a cubic lattice constant of a = $3.84\,\text{\AA}$. The shifted peak positions yield a lattice constant value of a = $3.85\,\text{\AA}$ (x = 0.17) and a = $3.86\,\text{\AA}$ (x = 0.28), respectively. The triangular configuration of the Mn spins is usually stable over a small compositional change in hexagonal members of the family \cite{khadka2020kondo,higo2018anomalous}. 
The thickness of the films is determined from the calibrated deposition rates and is confirmed from  XRR and AFM measurements. The XRR profiles shown in {\textcolor{blue}{Fig \ref*{fig:break1}(d)}} are acquired on samples with x= 0.09 and 0.17. The red curve represents the simulated model for $Mn_{3}Pt$/Si(100). Based on the fitting parameters, we obtain a thickness of 18 nm (0.09) and 20 nm (0.17) for the $Mn_{3}Pt$ layers, respectively. The  estimated roughness values of the $Mn_{3}Pt$ surface is 0.8 nm (0.09) and 1.2 nm (0.17), respectively. AFM topographs  ($20 \times 20$ \,\textmu m\textsuperscript{2}) presented in {\textcolor{blue}{Fig \ref*{fig:break1}(e--g)}} reveal the evolution of surface morphology with increasing Mn concentration. The root mean square roughness  obtained from the AFM topographs is less than 1 nm, consistent with the XRR analysis. The surface morphology of the samples is further investigated using STM, as shown in \hyperref[fig:break2]{\textcolor{blue}{Fig \ref*{fig:break2}}}.
The high-resolution STM topographs ($20 \times 20$ nm\textsuperscript{2}) confirm a well defined termination along (111) axis. Surface topography of films with x = 0.09 (\hyperref[fig:break2]{\textcolor{blue}{Fig \ref*{fig:break2}(a)}}) consists of islands and depression belonging to the consecutive surface terminations along the (111) axis. The height of the island and depression are determined from the line profile in \hyperref[fig:break2]{\textcolor{blue}{Fig \ref*{fig:break2}(b)}}. The peak height is $2.25\,\text{\AA}$, and the hole depths are approximately $2.5\,\text{\AA}$, close to the inter-planar spacing of the (111) planes.
 This also confirms the layer by layer mode of growth along the Kagome plane (111). STM topograph obtained on Mn rich films growth with (x = 0.17, 0.28) show similar atomically flat surface, as shown in \hyperref[fig:break2]{\textcolor{blue}{Fig.~\ref*{fig:break2}(c--d)}}. This also confirms that higher Mn flux rates don't induce any local variation in composition through defects or vacancies, as excessive Mn atoms occupy the Pt sites. 
 Extended doping of Mn atoms with Pt sites is also expected to tune the spin orientation in the Kagome plane as the energy difference between $T_{2}$ \& $T_{1}$ spin ordering arises from the spin-orbit coupling \cite{zemen2017frustrated}. The degeneracy of the two spin states is broken through compositional variation of Mn and Pt in this films.
 \begin{figure*}
    \centering
    \includegraphics[width=1.0\textwidth]{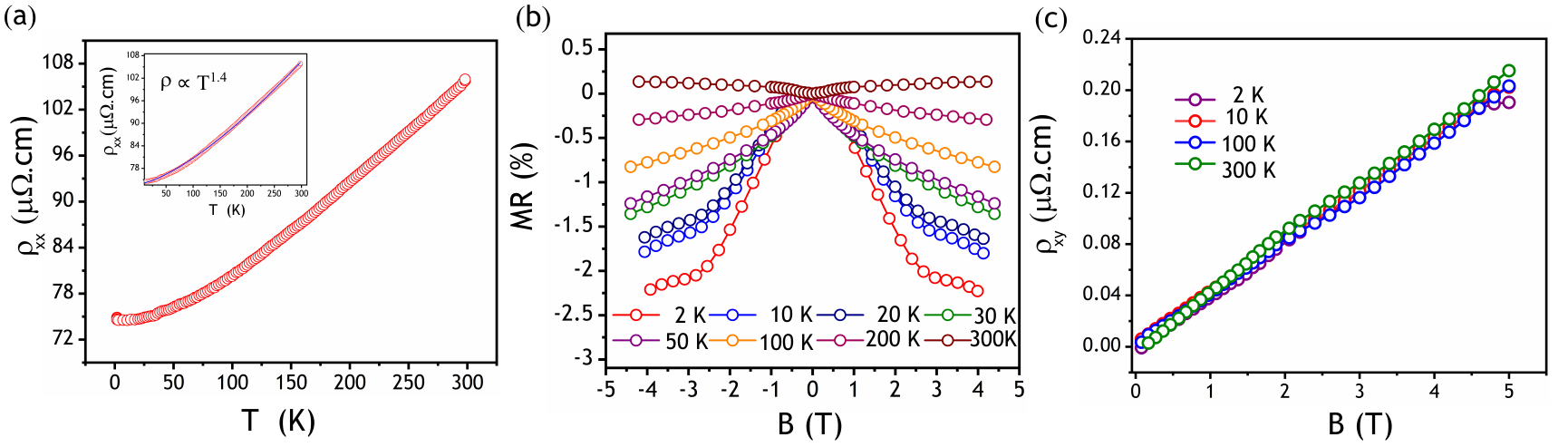} 
    \caption{ Magneto-transport studies on $Mn_{3+x}Pt_{1-x}$ thin films with x = 0.09. (a) Temperature dependent longitudinal resistivity ($\rho_{xx}$-T) measured in heating mode. The top inset in (a) shows the fitting between (50-300 K) using the power law equation : $\rho_{xx} = \rho_{xx0} + bT^n$. The observed value of n is 1.4. (b) Temperature-dependent transverse magneto-resistance (MR) measured between $\pm4~T$. The saturation in MR near $\pm3.5~T$ can be attributed to the transition from a  spin glass state to FM order. MR shows non-saturated behavior beyond 50 K due to negligible out-of plane moment in $T_{2}$ phase. (c) Field dependent Hall resistivity ($\rho_{xy}$) measured at four representative temperatures. The vanishing AHE at x = 0.09 is due to the non-topological spin state of $T_{2}$ phase.}
    \label{fig:break4}
\end{figure*}
 \section{\label{sec:level2}Magnetic measurement}
To distinguish between $T_{1}$ and $T_{2}$ phases, we study the temperature variation of magnetization on $Mn_{3+x}Pt_{1-x}$ thin films for x = (0.09, 0.17, 0.28). The measurements are conducted in zero field cool (ZFC) and field cooled (FC) modes with the magnetic field (B) normal to the film surface. In the ZFC mode, the sample is initially cooled to 2 K and the data is acquired during heating. During FC, the samples are precooled in the presence of a magnetic field.
The M-T curve in  \hyperref[fig:break3]{\textcolor{blue}{Fig \ref*{fig:break3}(a)}} is measured on a sample with x = 0.09. The ZFC-FC magnetization path shows an abrupt rise in moment in the low-temperature regime, indicating the presence of a  large spontaneous magnetization in the basal plane. This magnetic transition can be identified with the $T_{2}$ configuration and has been observed previously in Mn$_{3}$Sn bulk single crystals \cite{sung2018magnetic}. The sudden change in moment is associated with a transition from a spin glass state to a spiral AFM state. Below 30 K, the Mn spins develop a slight tilting from the basal plane and point towards the c-axis \cite{rout2019field,gibson2023magnetic}.
The isothermal magnetization (M-H) loops measured on the same sample is shown in \hyperref[fig:break3]{\textcolor{blue}{Fig \ref*{fig:break3}(d)}}. The M-H loops are obtained after the linear subtraction of the diamagnetic background of the substrate. The isotherm measured at 2 K exhibits a large coercive field of 1.75 T and reaches a saturated magnetic state near $H_{S}$ = 4 T. The isotherms demonstrate similar behavior until 30 K, attributed to the uncompensated moment from the sudden spin tilting. Above 50 K, the isotherms exhibit negligible hysteresis, indicating the presence of non-collinear AFM order. The saturation magnetization $M_{S}$ decreases by approximately 10 times from 25 $emu/cm^{3}$ to 2.5 $emu/cm^{3}$. M-T measurements for (x = 0.17) are shown in \hyperref[fig:break3]{\textcolor{blue}{Fig \ref*{fig:break3}(b)}}. The ZFC-FC curves maintain their reversibility between 2-300 K. This indicates that the triangular spin order is restricted to the basal plane down to low temperatures as observed for $Mn_{3}Ge$ bulk single crystals \cite{chen2021anomalous,qian2014exchange}.  
Previous reports on bulk $Mn_{3}Sn$ also indicate the absence of intermediate spiral AFM states in Mn-rich samples \cite{xia2024giant}.  \hyperref[fig:break3]{\textcolor{blue}{Fig \ref*{fig:break3}(e)}} shows the M-H loops  measured on the same sample (x = 0.17). The M-H loops show negligible hysteresis at all the measured temperatures. The small decline in saturation magnetization ($M_{S}$) with increasing temperature confirms the stabilization of chiral AFM order. The saturation moment ($M_{S}$) at 300 K increases to 8.7 $emu/cm^{3}$  due to a higher Mn concentration \cite{zhao2021magnetic}. The weak ferromagnetic behavior is a result of uncompensated Mn spins in the basal plane \cite{markou2018noncollinear}. The magnetometry results validate the presence of two non-collinear spin states grown under different conditions. The ZFC-FC curves measured on samples with (x = 0.28) also follow a reversible magnetization path as shown in \hyperref[fig:break3]{\textcolor{blue}{Fig \ref*{fig:break3}(c)}}. The non-collinearity between the Mn spins is further confirmed by the neglegible hysteresis observed in the M-H loops in \hyperref[fig:break3]{\textcolor{blue}{Fig \ref*{fig:break3}(f)}}.
A saturation magnetization of $M_{S} =15~emu/cm^{3}$ is observed at 300 K for x = 0.28. This indicates a change in magnetic phase from an incommensurate spiral  state ($T_{2}$) to a triangular spin configuration with an increase in Mn composition. $T_{2}$ is  stable at x = 0.09 and is recognizable as a non-topological intermediate state due to its vanishing AHE \cite{bai2021control}. The other phase at (x = 0.17, 0.28)  could possibly be the  topological $T_{1}$ phase with non-vanishing AHE, as predicted theoretically \cite{liu2018electrical}.

 \begin{figure*}
    \centering
    \includegraphics[width=01.0\textwidth]{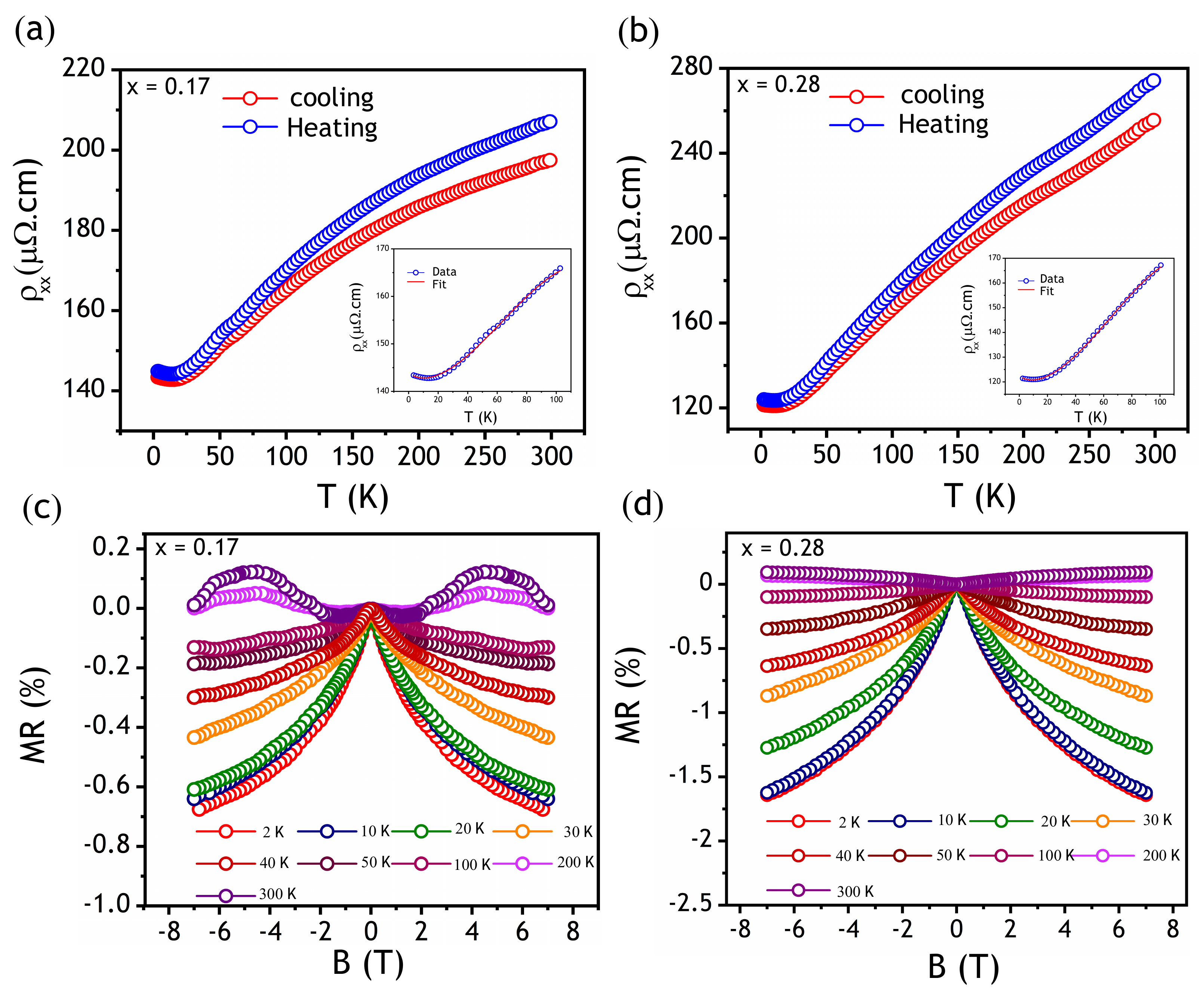} 
    \caption{Magneto-transport studies on $Mn_{3+x}Pt_{1-x}$ thin films with $x = 0.17, 0.28$. Panels (a) and (b) present the temperature-dependent longitudinal resistivity ($\rho_{xx}$) for $x = 0.17$ and $x = 0.28$, respectively. The insets illustrate the fitting of the low-temperature upturn using Eq. (1). Transverse magnetoresistance measurements at various temperatures are shown in (c) for $x = 0.17$ and in (d) for $x = 0.28$. A transition from negative magnetoresistance (NMR) to positive magnetoresistance (PMR) is observed near 200 K, with NMR arising from the suppression of low-energy magnetic fluctuations under increasing magnetic fields.}
    \label{fig:break5}
\end{figure*}

\section{\label{sec:level2}Magneto-transport measurements}
To confirm the topological signatures of the grown $Mn_{3+x}Pt_{1-x}$ thin films, we perform a detailed magneto-transport study for all three batches of samples (x = 0.09, 0.17, 0.28). Our preliminary transport measurements on polycrystalline samples are consistent with the previous reports and have been discussed in the supplementary section (Fig. S1) \cite{mukherjee2021sign}. 
We measure the temperature variation of electrical resistivity ($\rho_{xx}$-T) on films identified  with $T_{2}$ spin structure. \hyperref[fig:break4]{\textcolor{blue}{Fig \ref*{fig:break4}(a)}} shows the metallic response of $\rho_{xx}$  between 2 and 300 K with a a residual resistivity ratio (RRR = $\rho_{xx}$(300 K) / $\rho_{xx}$(2 K)) of 1.6. The inset shows the power law fitting with a $T^{1.4}$ dependence between 50-300 K. 
Field variation of magnetoresistance (MR) in \hyperref[fig:break3]{\textcolor{blue}{Fig \ref*{fig:break4}(b)}} show a steep decrease in MR attributed to the presence of a metastable spin glass phase, driven by an out-of-plane magnetic field. The change in magnetoresistance is evaluated using the following formula: $ MR = \left( \frac{\rho_{xx}(B) - \rho_{xx}(0)}{\rho_{xx}(0)} \times 100 \right)$. $\rho_{xx}(B)$ is the value of longitudinal resistivity at the magnetic field (B). The negative MR reaches a state of saturation at lower temperatures until Mn spins transition to an FM order \cite{zhu2020negative}. The steep decline in MR vanishes at 50 K as the spin structure transitions into a stable helical AFM order. These electrical signatures are consistent with the magnetic measurements shown in \hyperref[fig:break3]{\textcolor{blue}{Fig \ref*{fig:break3}(a)}} and \hyperref[fig:break3]{\textcolor{blue}{Fig \ref*{fig:break3}(d)}}.
Next, we measure isothermal Hall resistance  at different temperatures on the same sample. The magnetic field is along the (111) direction, perpendicular to the surface of the films. The contacts are made in Hall geometry using the standard four-probe method. To remove any contribution from the probe misalignment, Hall resistivity is anti-symmetrized over positive and negative fields using the following formula: ($\frac{\rho_{xy}{(+B)} - \rho_{xy}{(-B)}}{2}$). \hyperref[fig:break4]{\textcolor{blue}{Fig \ref*{fig:break4}(c)}} shows a  linear response of Hall resistivity, with  no contribution from the anomalous Hall effect (AHE). Traditionally, anomalous Hall effects (AHE) are strongly correlated  with magnetism, and both exhibit hysteresis under field variation. At x = 0.09, M-H loops exhibit large hysteresis below 50 K due to its out-plane component. The linear behavior of $\rho_{xy}$ indicates that AHE effect in $Mn_{3+x}Pt_{1-x}$ films is intrinsic and not driven by uncompensated magnetism. 
The scattering mechanisms are significantly different for  higher Mn concentrations (x = 0.17, 0.28), as evident from the ($\rho_{xx}$-T ) curves in \hyperref[fig:break5]{\textcolor{blue}{Fig \ref*{fig:break5}(a)}} and \hyperref[fig:break5]{\textcolor{blue}{Fig \ref*{fig:break5}(b)}}. The films exhibit similar metallic behavior with RRR values of 1.4 and 1.7 for $x=0.17$ and $x=0.28$, respectively. The low-temperature upturns are now more significant as compared to $T_{2}$ and have been fitted  into the following equation: \begin{equation}
\rho_{xx} = \rho_{xx0} + \alpha T^{2} + \beta T^{3} + \gamma T^{1/2}
\label{eq:eqn(1)}
\end{equation}
Where, $\alpha$, $\beta$, and $\gamma$ are the coefficients corresponding to electron-electron scattering (e-e), unconventional one-magnon scattering (1MS), and weak localization (WL) respectively \cite{wu2022structural}. The estimated parameters are summarized in table \ref{tab:mytable}.
\begin{table}[h!]
\centering
\caption{Fitting results of Equation (1)}
\label{tab:mytable}
\begin{tabular}{|c|c|c|c|c|}
\hline
Mn  & $\rho_{xx0}$ & $\alpha$ & $\beta$ & $\gamma$ \\ 
        ($x$) & [\textmu$\Omega$ cm] & [\textmu$\Omega$ cm K\textsuperscript{2}] & [\textmu$\Omega$ cm K\textsuperscript{3}] & [\textmu$\Omega$ cm K\textsuperscript{1/2}] \\ 
\hline
0.17 & 123.2 & 0.0112 & 0.00006 & -0.91 \\ 
\hline
0.28 & 142.1 & 0.008 & 0.00056 & -0.75 \\ 
\hline
\end{tabular}
\end{table}
By comparing the relative intensities of $\alpha$, $\beta$, and $\gamma$, it becomes evident that weak localization dominates the scattering mechanism near the low temperature. 

\begin{figure*}
    \centering
    \includegraphics[width=1.0\textwidth]{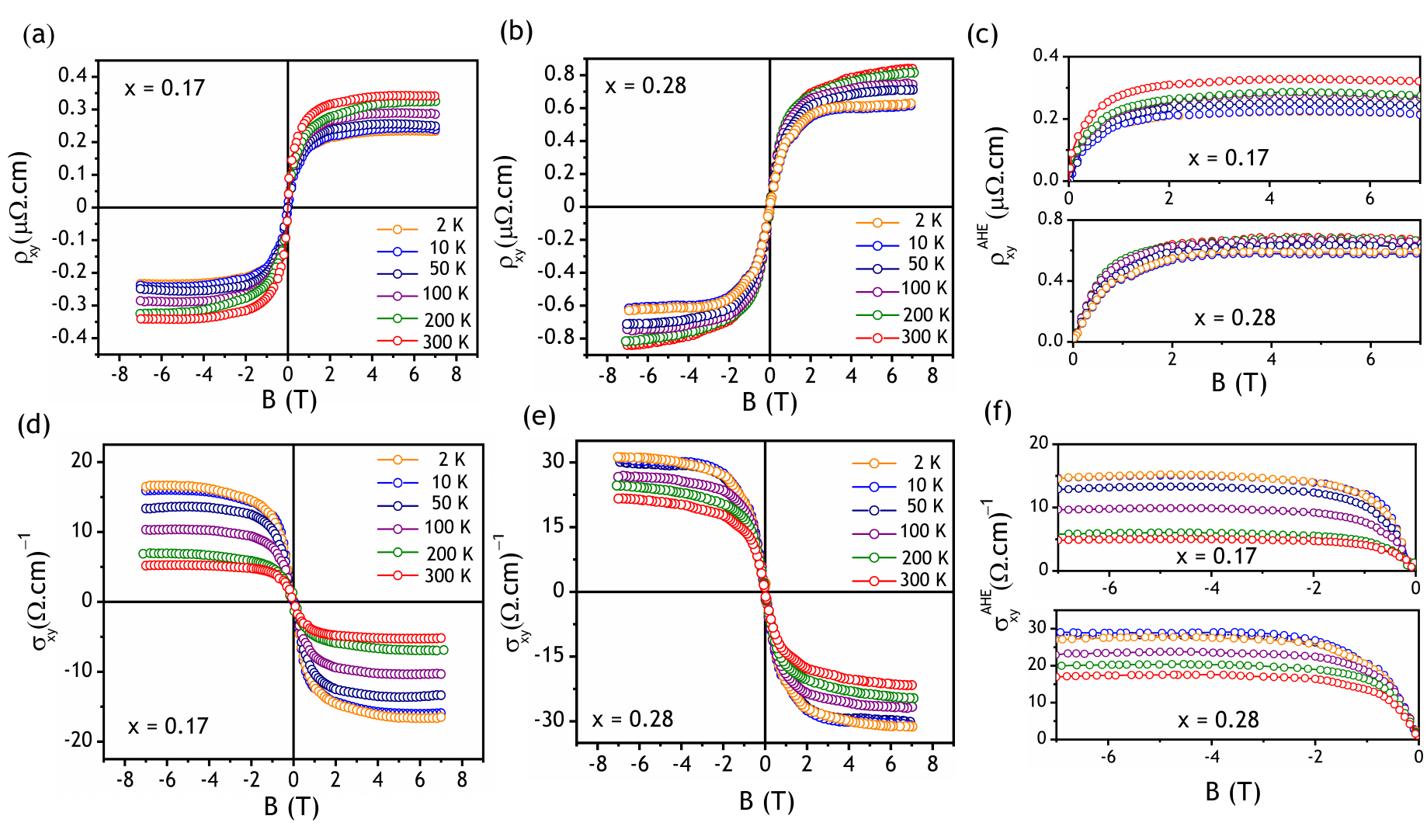} 
    \caption{ Field-dependent transverse resistivity and conductivity measurements for $x = 0.17$ and $x = 0.28$. Panels (a) and (b) present the isothermal variation of Hall resistivity ($\rho_{xy}$) measured within $B = \pm 7~T$ for $x = 0.17$ and $x = 0.28$, respectively. The non-linear variation of $\rho_{xy}$ confirms the stability of the topological $T_{1}$ spin state in Mn-rich compositions. (c) {$\rho_{xy}^{AHE}$}-B sweeps are estimated by subtracting the linear part of the Hall voltage in the high magnetic-field region. Panels (d) and (e) illustrate the variation of Hall conductivity ($\sigma_{xy}$) with the magnetic field for x= 0.17 and x =0.28. (f) {$\sigma_{xy}^{AHE}$}-B sweeps are estimated by subtracting the linear part of the Hall voltage in the high magnetic field range.
    }
    \label{fig:break6}
\end{figure*}
   
\begin{figure*}
    \centering
    \includegraphics[width=01.0\textwidth]{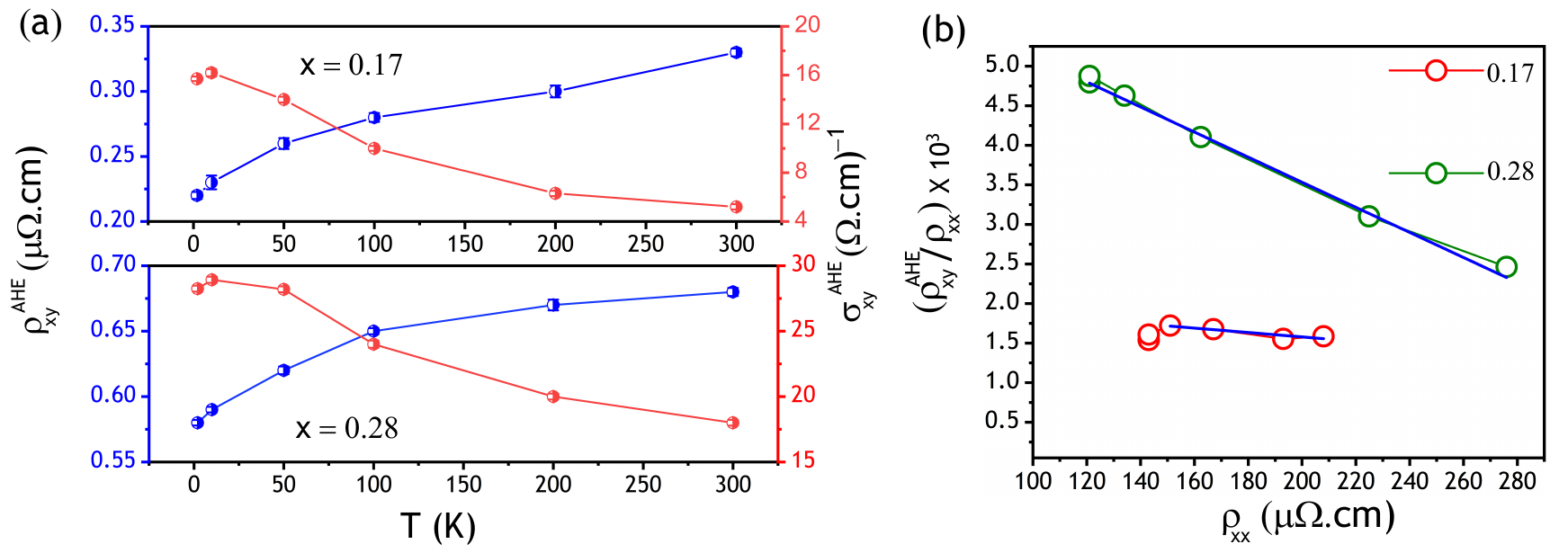} 
    \caption{(a) Variation of {$\rho_{xy}^{AHE}$} (blue) and {$\sigma_{xy}^{AHE}$} (red) with temperature is shown in the top (0.17) and bottom (0.28) respectively. (b) ($\rho_{xy}^{AHE}/\rho_{xx}$) plotted as a function of  $\rho_{xx}$  for x =0.17 (red) and x 
    =0.28 (green). The plot shows the scaling law for chiral $Mn_{3}Pt$ films with $T_{1}$ phase. The slope and intercept of the linear films represent the intrinsic and extrinsic contributions to anomalous Hall effect.}
    \label{fig:break7}
\end{figure*}
The magnetic field dependence of longitudinal resistivity ($\rho_{xx}$) for x = (0.17, 0.28) is shown in \hyperref[fig:break5]{\textcolor{blue}{Fig \ref*{fig:break5}(c-d)}}.
Temperature-dependent MR studies show a crossover between positive and negative MR near 200 K in both samples. The negative MR shows a saturating behavior at higher fields. This indicates that the spin scattering is suppressed, as the system is forced into a saturated state under external magnetic fields \cite{samanta2024emergence, roychowdhury2024enhancement}. The transition from negative to positive MR is attributed to the opposing effects of the two magnetic sublattices \cite{yamada1973magnetoresistance}. The change in MR is lower compared to the ($T_{2}$) phase, as the Mn spins are restricted to the (111) plane. Both magnetic and electrical measurements confirm  a non-collinear spin order that is persistent over the entire temperature range. To confirm  $Mn_{3+x}Pt_{1-x}$ is topological for (x =0.17,0.28), we measure the isothermal Hall resistivity of these samples. \hyperref[fig:break6]{\textcolor{blue}{Fig \ref*{fig:break6}(a--b)}} shows the field-dependent Hall resistivity ($\rho_{xy}(B)$) for x=(0.17 and 0.28), respectively. $\rho_{xy}(B)$ shows a sudden rise in the zero field region before saturating at higher fields, confirming the presence of AHE. The total Hall resistivity is expressed in the following equation \cite{pugh1953hall}:
\begin{equation}
\rho_{xy} = R_0 B + R_S M
\label{eq:eqn(2)}
\end{equation}
The first component ($R_{0}B$) is the ordinary Hall effect induced by the Lorentz force and the second component represents the contribution of the anomalous Hall term ($\rho_{xy}^{AHE}$). The positive slope observed in the high-field regime suggests the dominance of hole pockets near the Fermi level. The non-linear behavior is suggestive of a dominating anomalous Hall term for both values of x. $R_{0}$ and $\rho_{xy}^{AHE}$ at each  temperature is extracted from a linear fit to the  $\rho_{xy}$-B data as shown in the supplementary section (Fig. S2). $R_{0}$ and $\rho_{xy}^{AHE}$ are evaluated from the slope and intercept, respectively. The carrier concentration evaluated from ($R_{0} = 1/ne$) is found to vary between ($10^{30}$$cm^{-3}$-$10^{31}$$cm^{-3}$) as the Mn composition increases from 0.17 to 0.28. The variation in carrier concentration with temperature is shown in the supplementary section (Fig. S2). $\rho_{xy}^{AHE}$-B plots in \hyperref[fig:break6]{\textcolor{blue}{Fig. \ref*{fig:break6}(c)}} are obtained by subtracting the linear component from the high-field region. This indicates that the Mn spins adopt a triangular spin order ($T_{1}$), leading to such robust AHE in this batch of samples.
 The  transverse hall conductivity is estimated  using the  following equation :
 \begin{equation}
\sigma_{xy} = -\frac{\rho_{xy}}{(\rho_{xx}^2 + \rho_{yx}^2)}
\label{eq:eqn(3)}
\end{equation}
The calculated Hall conductivity ($\sigma_{xy}$ ) is plotted as a function of magnetic field (B) in \hyperref[fig:break6]{\textcolor{blue}{Fig \ref*{fig:break6}(d)}} and \hyperref[fig:break6]{\textcolor{blue}{Fig \ref*{fig:break6}(e)}}. $\sigma^{AHE}_{xy}$ is evaluated by extrapolating  $\sigma_{xy}$ from the high field region and  has been demonstrated in the supplementary section (Fig. S2). $\sigma^{AHE}_{xy}$ (2 K) increases from  16 $\Omega^{-1} \mathrm{cm^{-1}}$ (x = 0.17) to 28 $\Omega^{-1} \mathrm{cm^{-1}}$ (x = 0.28).  
 The temperature variation of $\sigma^{AHE}_{xy}$ and $\rho^{AHE}_{xy}$ is shown in \hyperref[fig:break7]{\textcolor{blue}{Fig \ref*{fig:break7}(a)}}. A comparative study show an increase in  $\rho_{xy}^{AHE}$ with increasing Mn concentration. The observed values of $\rho_{xy}^{AHE}$ at 300 K are approximately 0.32 ($\mu\Omega.cm$) and 0.68 ($\mu\Omega.cm$) for x = 0.17 and 0.28 respectively. 
The Anomalous Hall  resistivity and conductivity  have intrinsic or extrinsic origins. It can be mathematically expressed {\textcolor{blue}{equation (4)}}:
\begin{equation}
 \rho_{xy}^{AHE}/\rho_{xx} = a_{sk}  + \sigma_{in} \rho_{xx}
 \label{eq:eqn(4)}
\end{equation}
The first term indicates contributions from extrinsic mechanisms like skew scattering, while the second term  is the intrinsic AHC \cite{nagaosa2010anomalous,xiao2010berry}, which is predominantly driven by the non-vanishing Berry curvature. The intrinsic contribution ($\sigma_{in}$) of the AHC  can be determined from the slope in \hyperref[eq:eqn(3)]{\textcolor{blue}{equation (4)}}.
\hyperref[fig:break6]{\textcolor{blue}{Fig.~\ref*{fig:break6}(c)}} shows the linear fitting of $\rho_{xy}^{AHE}/\rho_{xx}$ with $\rho_{xx}$ for both concentrations of $Mn_3Pt$. Based on the fitting parameters, we obtained $\sigma_{in}$ values of approximately 4.74 $\Omega^{-1} \mathrm{cm^{-1}}$ and 12 $\Omega^{-1} \mathrm{cm^{-1}}$ for $x = 0.17$ and $x = 0.28$, respectively. Increase in Pt concentration enhances the extrinsic contribution as observed for Ir doped MnPtSn alloys \cite{jamaluddin2022extrinsic}. This could be attributed to the slight projection of the Mn spins relative to the (111) plane due to high SOC of Pt \cite{chen2014anomalous}.
The skew scattering contribution ($a_{sk}$) is neglegible for $Mn_3Ge$ and $Mn_3Sn$ but in cubic crystals both vector scalarity (VSC)-induced intrinsic AHE and scalar spin chirality (SSC)-induced skew scattering contribute to the AHE \cite{ding2021field,ishizuka2018spin}. The $a_{sk}$ values are of the order $10^{-3}$ in both the cases, indicating contribution from  scalar spin chirality (SSC)-induced skew scattering effects in the cubic family $Mn_{3}X$ (X= Ir, Pt) \cite{xu2024universal}. 

\section{\label{sec:level2}Electronic Structure Calculations}

\begin{figure}
    \centering
    \includegraphics[width=0.470\textwidth]{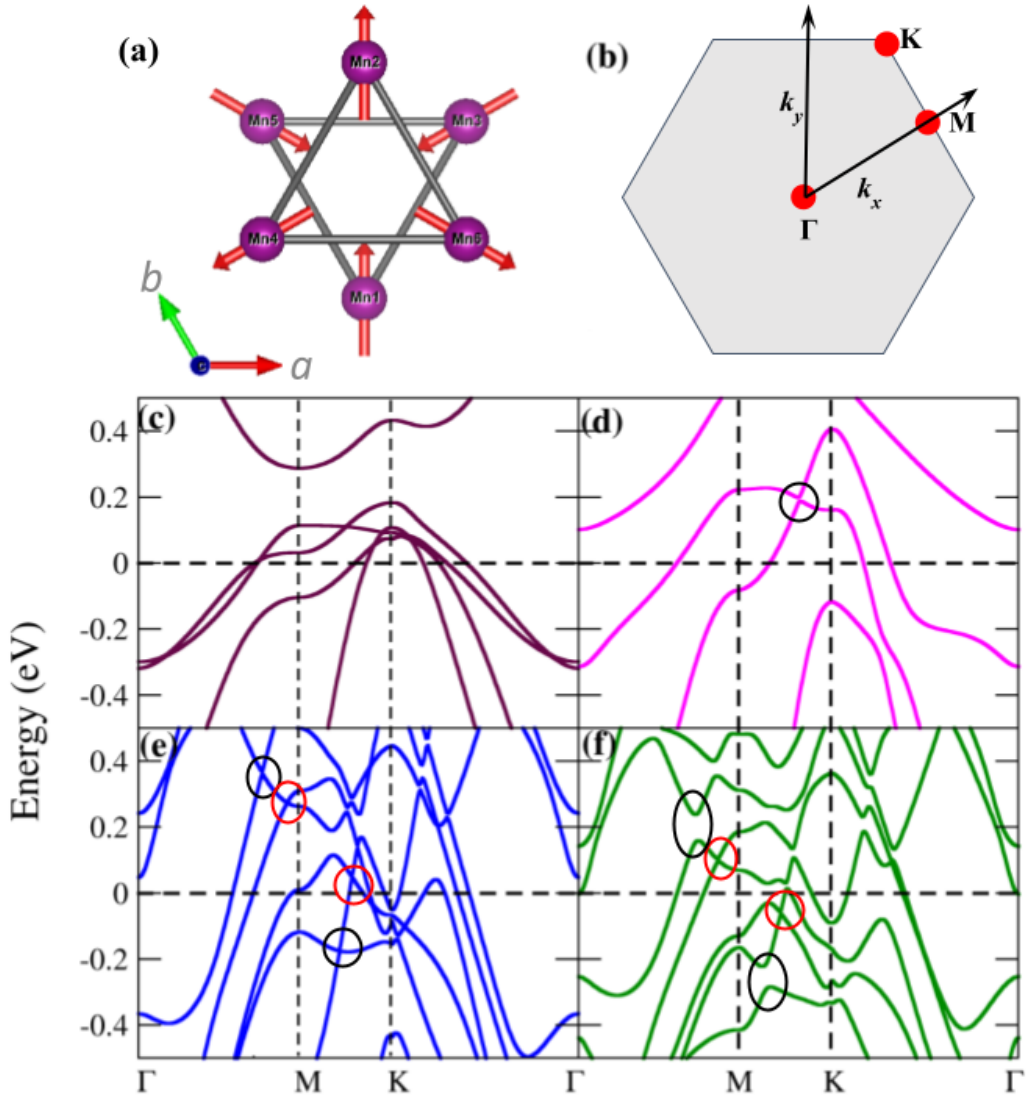} 
    \caption{ (a) The Mn only sublattice in the (111) hexagonal plane showing the spin arrangement of the T$_1$ configuration. (b) The two dimenssional hexagonal Brillouin zone (BZ) with relevant high symmetry points marked. Band structure plot for collinear AFM along the high symmetry k-points calculated within (c) GGA+$U$ and (d)  GGA+$U$+SOC scheme. Band structure plot for non-collinear T$_1$ AFM configuration along the high symmetry k-points calculated within (e) GGA+$U$ and (f)  GGA+$U$+SOC scheme.}    
    \label{fig:theory}
\end{figure}
To understand the microscopic details of the (111) grown thin film of Mn$_3$Pt, we have done electronic structure investigation of the hexagonal layered Mn$_3$Pt, where three Mn sites are in a plane of (111) form a Kagome lattice and two such planes are stacked along the perpendicular to that of the (111) direction. The symmetry analysis of such (111) oriented Mn$_3$Pt film  which is technically equivalent to that of the hexagonal film grown along the crystallographic $c$ direction. In that scenario the structure possess combined $\mathcal{M}$ plane along (111) and non-integer translation perpendicular to the $\mathcal{M}$ plane combined preserves which leads to non-symorphic glide symmetry in the grown thin film. The experimentally determined all-in and all-out non-collinear spin configuration named as T$_1$ shown in Fig \ref*{fig:theory}(a) and the total energy calculations show that it is energetically $\approx$ 1.29 eV/f.u lower compared to that of the T$_2$ spin configuration as shown in Fig \ref*{fig:break1}(a), consistent with the experimental findings. To understand the role of the specific T$_1$ type non-collinear spin configuration and SOC in deriving the topological non-trivial phase, we have calculated the band structures without and with SOC for both collinear and non-collinear AFM spin configurations as shown in Fig \ref*{fig:theory}(c)-(f). The calculated band structure for collinear AFM spin configuration in absence of SOC (Fig \ref*{fig:theory}(c)) clearly shows the metallic nature of electronic state with Mn-$d$ bands crossing the Fermi energy throughout the BZ. Moreover there are some band crossing like signature along the $\Gamma-M-K$ path. After switching on the SOC in the collinear spin arrangement, we found two doubly degenerate bands crosses the Fermi energy, however some accidental band crossing has been removed, while $M-K$ band crossing intact. Interesting things happen when we incorporate the non-collinear AFM spin configuration in the form of T$_1$. In the absence of SOC (Fig \ref*{fig:theory}(e)), there are multiple linear band crossing occurs marked by the circles, close to the Fermi energy. Two such linear band crossings along the $\Gamma-M$ direction and $M-K$ direction respectively was observed. The metalicity has also be more pronounced as many bands crosses the Fermi energy. Point to be noted that out of such four band crossings two of them close to Fermi energy (marked by red circles) and two of them are away from Fermi energy (marked by black circles). Now if we switch on the SOC in the non-collinear T$_1$ type AFM state, as shown in Fig \ref*{fig:theory}(f), some of the crossing points are open up gap (marked by black circles in  Fig \ref*{fig:theory}(f)) due to the breaking of degeneracy in presence of SOC. However two band crossing marked by red circles in Fig \ref*{fig:theory}(f) still persist very close to Fermi energy. The crossing in the conduction band is $\approx$ 100 meV the Fermi energy along the $\Gamma-M$ direction, while in the valence band the crossing is at $\approx$ 50 meV below the Fermi energy along the $M-K$ direction in the BZ. Interesting fact is that above stated these two band crossing appears in the non-high symmetry points even in the presence of SOC. The band crossing is protected even in the presence of SOC due to the presence of nonsymorphic symmetry involved in the structure \cite {schoop2016dirac, bian2016topological} and in this case it is the glide symmetry that protect the Dirac band crossing even in the presence of SOC. The topological non-trivial Dirac crossing in the bulk band drives the non-vanishing Berry curvature in the system which in turn responsible for the signature of the intrinsic AHE in the experiments. Moreover, comparing the Fig \ref*{fig:theory}(e) and (f), we can also understand that strength of SOC also plays crucial role in developing the topological non-trivial band crossings, as it is very evident that some of the band crossing that appears in the case of without SOC, that gets washed out in presence of SOC as a results decreases the number of the linear crossing in the entire BZ.

\section{\label{sec:level2}Conclusions}
In summary, we synthesized cubic $Mn_3$Pt(111) thin films on Si(100) using magnetron sputtering technique and examined how variation of Mn composition affect their micro-structural, magnetic and magneto-transport characteristics. Combined X-ray reflectivity and high resolution STM measurements confirms that the $Mn_3$Pt(111) films grow epitaxially with a layer-by-layer fashion along the Kagome plane. We observed the presence of two distinct non-collinear spin configurations ( $T_1$) and ($T_2$), which vary with the Mn compositional ratio. The $T_2$ phase remains stable in the un-doped state ($x = 0.09$) while the $T_1$ phase stabilizes in Mn-rich samples ($x = 0.17$  $0.28$). Electronic structure calculations identified the non-trivial Dirac crossing in the bulk band structure protected by the non-symorphic glide symmetry in the case of the T$_1$ type non-collinear spin configuration which is responsible for the observed AHE up to 300 K. The observed large AHE is primarily driven by the intrinsic Berry curvature of the momentum space due to the presence of the Dirac state around 50 and 100 meV below and above the Fermi energy respectively. The scaling behavior of the ratio $\rho_{xy}^{AHE}/\rho_{xx}$ versus $\rho_{xx}$ demonstrates a linear dependence, suggesting that Berry curvature dominates over skew scattering mechanisms. Our present investigation reveals the mechanism of the chemical tuning via optimal Mn doping to stabilize desired non-collinear spin texture which can lead to a topological non-trivial band structure and Berry curvature driven room temperature intrinsic AHE. The observation of spontaneous AHE establishes $Mn_3Pt$ as a potential candidate for topologically driven spintronic devices and applications.

\section{\label{sec:level2}Acknowledgment}
SM acknowledge the financial grant support from SERB Core Research Grant (CRG/2023/008193). IS and SM acknowledge partial support from the DST-funded project 'CONCEPTS' under the Nanomission program (DST/NM/QM-10/2019), Government of India. SM and IS thanks Vikas Chahar and Ratnamala Chatterjee for fruitful discussion on UHV sputtering system. Thin films characterizations (XRD, AFM) and magnetization measurements (MPMS) are performed using IIT Delhi Physics department shared facility. Authors would also like to acknowledge CRF and NRF, IIT Delhi for PPMS and EDX facility.

\bibliography{bibliography}

\end{document}